\documentclass[pre,10pt,twocolumn,groupedaddress,showpacs,floatfix,superscriptaddress]{revtex4}
\usepackage{graphicx}
\usepackage{ulem}
\usepackage{dcolumn}
\usepackage{amsmath}
\usepackage{bm}

\begin{document}
\title{Young's modulus of polyelectrolyte multilayers from microcapsule 
swelling }

\author{O. I. Vinogradova}
\email[Corresponding author: ]{vinograd@mpip-mainz.mpg.de}
\affiliation{Max Planck Institute for Polymer Research, Ackermannweg 10, 55128 Mainz, Germany}
\affiliation{Laboratory of Physical Chemistry of Modified Surfaces, Institute of Physical Chemistry, Russian Academy of Sciences, 31 Leninsky Prospect, 119991 Moscow, Russia}

\author{D. Andrienko}
\affiliation{Max Planck Institute for Polymer Research, Ackermannweg 10, 55128 Mainz, Germany}

\author{V. V. Lulevich}
\affiliation{Max Planck Institute for Polymer Research, Ackermannweg 10, 55128 Mainz, Germany}
\affiliation{Laboratory of Physical Chemistry of Modified Surfaces, Institute of Physical 
Chemistry, Russian Academy of Sciences, 31 Leninsky Prospect, 119991 Moscow, 
Russia}

\author{S. Nordschild}
\affiliation{Max Planck Institute for Polymer Research, Ackermannweg 10, 55128 Mainz, Germany}

\author{G. B. Sukhorukov}
\affiliation{Max Planck Institute for Colloid and Interface Research, Golm 14424, Potsdam, Germany}

\date{\today}

\begin{abstract}
We measure Young's modulus of a free polyelectrolyte multilayer film by 
studying osmotically induced swelling of polyelectrolyte multilayer 
microcapsules filled with the polyelectrolyte solution. Different filling 
techniques and core templates were used for the capsule preparation. Varying 
the concentration of the polyelectrolyte inside the capsule, its radius and 
the shell thickness yielded an estimate of an upper limit for Young's 
modulus of the order of 100 MPa. This corresponds to an elastomer and 
reflects strong interactions between polyanions and polycations in the 
multilayer.
\end{abstract}
\pacs{46.70.De, 68.37.Ps, 81.05.Lg}
\maketitle

\section{Introduction}

Molecularly thin polyelectrolyte multilayer films are composed of 
alternating layers of oppositely charged polyions and are important for a 
variety of potential applications~\cite{1,2}. Supported multilayer films 
are normally produced via layer-by-layer (LbL) adsorption of polyanions and 
polycations on a planar~\cite{1} or spherical~\cite{3} charged solid 
surface. In the later case, the colloidal template can be dissolved to give 
so-called polyelectrolyte microcapsules~\cite{4}. The shell of such 
microcapsules is nothing more than a free standing multilayer film. The free 
thin film geometry allows investigation of properties not accessible in the 
bulk or in supported films and helps to gain a better understanding of 
polyelectrolytes in general. 

There have been a number of recent experimental studies of mechanical 
properties of thin multilayer films. First, by studying osmotically induced 
buckling of ``hollow'' (water inside) capsules immersed in a polyelectrolyte 
solution~\cite{5}, the Young's modulus was found to be above 1000 MPa, 
close to the elasticity of the \textit{bulk plastics}~\cite{6}. The second, more recent, 
approach is based on measuring the deformation of microcapsules under 
applied load using an atomic force microscope 
(AFM)~\cite{7,8,9}. This method yielded an estimate 
of the \underline {lower} limit for Young's modulus of the order of 1-10 MPa 
which corresponds to an \textit{elastomer}~\cite{6}. The reason for such a discrepancy 
between two approaches might be hidden in the assumptions of the models used 
to fit experimental data. The first model assumes that the shell is highly 
permeable for water, even on short timescales of the relaxation of buckling 
deformation. The second approach treats the shell as impermeable for water 
on short timescales, relying on the conservation of capsule volume for small 
deformations. The accuracy of existing experimental results does not allow 
to conclude which approach is more realistic.

In this paper we propose an alternative way to probe the elastic properties 
of a polyelectrolyte multilayer. The method is based on studying the 
swelling of microcapsules filled with a solution of a strong 
polyelectrolyte. The size of a swollen capsule depends on the Young's 
modulus of the capsule shell. Fitting the experimental data to the 
prediction of a simple model yields Young's modulus of the polyelectrolyte 
multilayer of the order of 100 MPa. We argue that our results give the 
\underline {upper} limit of Young's modulus. Taken together with recent AFM 
observations~\cite{7}, this result leads to the conclusion that the 
mechanical behavior of a polyelectrolyte multilayer is that of an 
\textit{elastomer.}

\section{Theoretical model}

We consider a ``filled'' (polyelectrolyte solution inside) capsule immersed 
in a low molecular weight solvent (water in our case). The capsule swells 
due to excess osmotic pressure of the inner solution. The osmotic pressure 
of polyelectrolyte solutions is the sum of polymer and counterion 
contributions. However, in a salt-free solution the latter exceeds the 
osmotic pressure due to polymer itself by several orders of 
magnitude~\cite{10,11,12}. The capsule shell is permeable to 
the solvent (on time scales larger than the characteristic diffusion time), 
but impermeable to the encapsulated polymer of high molecular weight. Then 
the solvent diffuses into the capsule until the elastic force of the 
stretched shell balances the osmotic pressure. We assume that the inner 
polyelectrolyte solution remains electroneutral, i.e. all counter ions due 
to polyelectrolyte dissociation remain in the capsule interior. A similar 
assumption was used in Ref.~\cite{5}. 

We also assume that, for small relative deformations, the response of the 
capsule shell is elastic. If the capsule swells from the initial radius $r_0 
$ to a final radius $r$ the energy of stretching of the shell is given by 
the elastic theory of membranes~\cite{13}

\begin{equation}
\label{eq1}
G = 4 \pi \frac{E}{1 - \nu }h\left( {r - r_0 } \right)^2,
\end{equation}
where $E$ is Young's modulus, $\nu $ is Poisson's ratio ($\nu \approx 1 / 
2)$, and $h$ is the shell thickness ($h < < r_0 )$. 

For a dilute solution of the inner polyelectrolyte, or, alternatively, small 
concentration $n = N / V$ of counterions in the capsule, the osmotic 
pressure induced by the counterions reads

\begin{equation}
\label{eq2}
\Pi = \varphi \frac{N}{V}k_B T.
\end{equation}
Here $N$ is the number of counterions and $\varphi \le 1$ is the osmotic 
coefficient, defined as the ratio of experimentally measured osmotic 
pressure $\Pi $ to the ideal osmotic pressure of all counterions. The 
difference between these two values is due to a fraction of condensed 
counterions being bound to the polyelectrolyte chain and not contributing to 
the osmotic pressure~\cite{14,15,16}. 

The work done by the osmotic pressure $\Pi $ to swell the capsule from 
radius $r_0 $ to $r$ then reads

\begin{equation}
\label{eq3}
A = \int_{V_0 }^V {\Pi dV} = - 3\varphi Nk_B T\ln \frac{r}{r_0 }.
\end{equation}

The equilibrium radius of the capsule is given by the minimum of the total 
energy $F = G + A$, where $\partial F / \partial r = 0$, giving

\begin{equation}
\label{eq4}
r = \frac{1}{2}r_0 \left( {1 + \sqrt {1 + \frac{3}{2}\frac{(1 - \nu )\varphi 
Nk_B T}{Ehr_0^2 }} } \right).
\end{equation}

To relate the number of counterions to the concentration of the polymer in 
the solution we need to know the degree of dissociation, i.e. the number of 
counterions per monomer. For a strong polyelectrolyte this number could be 
taken as 1, i.e. one counterion per monomer. Then

\begin{equation}
\label{eq5}
N = N_A cV_0 = \frac{4}{3}\pi r_0^3 cN_A ,
\end{equation}

\noindent
where $c$ is the concentration of the polymer solution in the capsule before 
it swells, $N_A $ is the Avogadro number. Substituting Eq. (\ref{eq5}) into Eq. (\ref{eq4}) one obtains 

\begin{equation}
\label{eq6}
r = \frac{1}{2}r_0 \left( {1 + \sqrt {1 + \frac{2\pi r_0 }{h}\frac{(1 - \nu 
)}{E}\varphi cRT} } \right),
\end{equation}

\noindent
where $R = k_B N_A $ is the universal gas constant. Equation (\ref{eq6}) relates the size of the swollen capsule to the concentration of the inner solution, 
thickness of the capsule shell, and Young's modulus. 

Our model allows design of a swelling experiment. To determine the Young's 
modulus, one can measure the deformation of the capsule as a function of 
concentration $c$ and the shell thickness $h$ and then fit the experimental 
data to Eq. (\ref{eq6}). 

\section{Experimental}

For an initial application of our approach we have chosen to study two types 
of capsules, characterized by different methods of encapsulation and 
templated on different cores. The first type are ``filled'' capsules 
prepared on manganese carbonate templates by a controlled 
precipitation~\cite{8,17}, i.e. by an assembly of the 
inner layer of polyelectrolyte shell by means of multivalent ions with the 
subsequent extraction of these ions and polymer release into the capsule 
interior. The second type of ``filled'' capsules were made from the 
``hollow'' ones, templated on melamine formaldehyde particles, by regulating 
their permeability for high molecular weight 
polymers~\cite{9,18}.

As a polyelectrolyte for encapsulation we have chosen sodium polystyrene 
sulfonate (PSS). The behavior of this highly charged flexible polyanion in 
salt-free conditions has been studied both theoretically~\cite{15} and 
experimentally~\cite{10,11,19,20}. The value of $\varphi 
$ for PSS was found to range from 0.2 to 1 and suggests some condensation of 
counter-ions~\cite{15,17}. 

\subsection{Materials and Methods}

\subsubsection{Materials}

The fluorescent dye Rhodamine B isothiocyanate (RBITC), monomers allylamide 
and 4-styrenesulfonic acid sodium salt hydrate (SS), shell-forming 
polyelectrolytes poly(sodium 4-styrenesulfonate) (PSS; $\rm M_{W} \sim 
70,000 \rm g/mol$  ) and poly(allylamine hydrochloride) (PAH; $ \rm M_{W} 
\sim  70,000 \rm g/mol$), and ethylenediaminetetraacetic acid (EDTA) were 
purchased from Sigma-Aldrich Chemie GmbH, Germany. Hydrochloric acid (HCl), 
acetone and sodium chloride (NaCl) were purchased from Riedel-de Ha\"{e}n; 
Germany. The ionic initiator potassium peroxodisulfate 
(K$_{2}$S$_{2}$O$_{8})$ for the radical polymerisation and Y(NO$_{3})_{3}$ 
were obtained from Merck GmbH; Germany. All chemicals were of analytical 
purity or higher quality and were used without further purification. 

To produce labelled PSS for encapsulation we used a modification of a method 
published in Ref.~\cite{21}. First, labelled allylamide was made which 
was afterwards mixed with 4-styrenesulfonic acid sodium salt hydrate and 
then copolymerised radically. The allylamide was mixed with RBITC solved in 
ethanol. The mixture was stirred for four hours at room temperature. 
Afterwards SS was added in an amount corresponding to a label grade of about 
200 monomer units each. Then K$_{2}$S$_{2}$O$_{8}$ was added to this 
solution as an ionic initiator for the radical polymerisation. The mixture 
was heated up to $80^0$C and was stirred for four hours in a nitrogen 
atmosphere. Afterwards it was centrifuged with a membrane which filters 
molecules with a molecular weight of 100,000 g/mol or more, and 
polydispersity was found to be M$_{W}$/M$_{N} \sim $1.8. The remaining PSS 
was chopped into small parts and washed with ethanol until no more colour 
could be observed in the filtrate. 

Suspensions of monodispersed weakly cross-linked melamine formaldehyde 
particles (MF-particles) with a radius of $r_0 = 2.0\pm 0.1 \quad \mu $m were 
purchased from Microparticles GmbH (Berlin, Germany). 

The manganese carbonate template (MnCO$_{3})$ was prepared by a mixing 
method described in Ref.~\cite{22}. Briefly, acidic manganese 
sulfate solution ($9\times 10^{ - 3}$~M, pH=4.2 adjusted by sulfuric acid) 
was added at 1:1 volume ratio to $2.25\times 10^{ - 3}$~M NH$_{4}$HCO$_{3}$. 
Then the stirred mixture was aged at $50^{0}$C for 16 hours. The resulting 
MnCO$_{3}$ particles had a spherical shape with a radius of 1.85$\pm $0.2 
$\mu $m and $2.5\pm 0.2\;\mu $m . 

Water used for all experiments was purified by a commercial Milli-Q Gradient 
A10 system containing ion exchange and charcoal stages, and had a 
resistivity higher than 18M$\Omega $/cm.

\subsubsection{Methods}

\subsubsection*{Capsule Preparation}

Two different approaches have been exploited for preparation of filled 
capsules. 

\begin{figure}[htbp]
\centerline{\includegraphics[width=8cm]{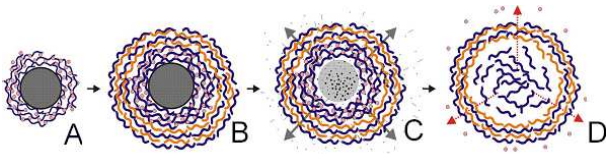}}
\caption{Scheme illustrating the preparation of ``filled'' microcapsules of the first type.
}
\label{fig1}
\end{figure}

\begin{figure}[htbp]
\centerline{\includegraphics[width=8cm]{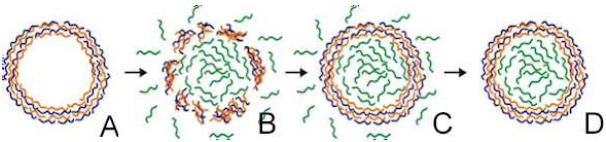}}
\caption{Scheme illustrating the preparation of ``filled'' microcapsules of the second type.}
\label{fig2}
\end{figure}

The preparation of filled capsules of the first type consisted of several 
steps (Fig.1). The first step (Fig.1a) was surface controlled precipitation 
of labeled PSS (by complex formation with Y$^{3 + }$ ions) on the surface of 
MnCO$_{3}$~\cite{8,23,24}. By varying the number of 
precipitated layers we were able to tune the surface density for adsorbed 
layers, and, therefore, the number of PSS molecules precipitated on MnCO$_{3 
}$ particles. In our experiments we prepared samples with 20-80 layers of 
adsorbed labeled PSS, which should lead to a surface density, $\rho $, in 
the range from $5\times 10^{ - 5}$ to $2\times 10^{ - 4}$mol/m$^{2}$. These 
values were calculated by assuming that the surface density for a monolayer 
is constant, and that the amount of adsorbed PSS grows linearly with the 
number of deposited layers ~\cite{17,24}. As a result, each 
capsule contained up to a few pg of PSS, depending on the template size and 
amount of assembled layers. With such a method, the concentration of 
encapsulated polyelectrolyte is approximately equal to $c = 3\rho / r_0 $. 
The MnCO$_{3}$ particles were dissolved in $ \sim $ 1mol/L HCl after 
assembling of 7 layers in order to facilitate the process of core removal 
since shells with a thickness of more then 10 layers prevent ion 
penetration~\cite{25}. Core dissolution led to the formation of 
``double shell'' structured capsules (Fig.1c). The inner shell formed by the 
PSS/Y$^{3 + }$ complex was not stable and was decomposed either by metal-ion 
complex agents (EDTA) or in salt solution. Yttrium ions were gradually 
expelled out of the outer stable shell formed by PSS/PAH while PSS molecules 
were released into the capsule interior (Fig.1d). Then the polyelectrolyte 
capsules were covered additionally with a number of layers varied in the 
interval from 1 to 13 to tune the final shell thickness, which, as a result, 
varied from 4 to 10 PSS/PAH bilayers. 

The ``filled'' capsules of the second type were made from pre-formed 
``hollow'' capsules~\cite{26}. The original ``hollow'' capsules were 
produced by a standard LbL assembly of 4 PSS/PAH bilayers on MF particles. 
MF-particles coated with PSS/PAH multilayers were dissolved in $ \sim $ 
1mol/L HCl and MF-oligomers were removed by washing, as described 
in~\cite{27}. Then the encapsulation of polymer included several 
steps. (Fig.2). The original ``hollow'' capsules (Fig.2a) were exposed to 
acetone/water mixture (1:1) to make the polyelectrolyte multilayer permeable 
for high molecular weight polymer~\cite{9}, and the PSS molecules 
were added to the mixture. The permeable state of the capsule shell allows 
the polymers to penetrate inside (Fig.2b). During the encapsulation process 
the PSS concentration was increased gradually to avoid an osmotic collapse 
of the microcapsules~\cite{5}. The initial concentration was 1g/L, and 
was doubled every hour. When the required concentration was reached, the 
mixture was diluted with water and the multilayer shells were assumed to 
return to an impermeable state (Fig.2c). After washing in pure water the 
capsules contain polymer solution (Fig.2d). With such a method, the 
concentration of encapsulated polyelectrolyte (before swelling) is 
approximately equal to the final concentration in the bulk.

\subsubsection{Confocal Laser Scanning Microscopy. }

To scan the capsule shape and to measure the concentration of PSS inside the 
capsules we used a commercial confocal microscope manufactured by Olympus 
(Japan) consisting of the confocal laser scanning-unit Olympus FV 300 in 
combination with an inverted microscope Olympus IX70 equipped with a high 
resolution 100\sout{ }oil immersion objective. The excitation wavelength was 
chosen according to the label Rhodamine (543 nm). The z-position scanning 
was done in steps of 0.2 $\mu $m. The diameters of the swollen capsules were 
determined optically with an accuracy of 0.4 $\mu $m. Concentration 
measurements were performed via the fluorescence intensity coming from the 
interior of the PSS containing capsules. In this case we assumed that 
fluorescence is directly proportional to PSS concentration and used a 
calibration curve of fluorescence intensity of free polymer in the bulk 
solution. The measured concentration was then recalculated to the initial 
concentration inside unswollen capsules. 

\section{Results and discussion}

The 3D confocal scanning showed that the capsules filled with PSS have 
spherical form. It should be noted that immediately after the preparation 
the capsule sizes were close to the size of the templates used for their 
preparation. The capsules swell for at least several days before reaching 
their equilibrium size, so that all the measurements of the radius of the 
swollen capsules were performed 2 weeks after filling with PSS. Fig.3 (top) 
shows a typical confocal fluorescence image of the swollen capsules. The 
fluorescence intensity suggests a uniform concentration in the capsule's 
interior. The bright interior of the capsules did not change with time, and 
there was no fluorescence signal from water. This proves that the capsules 
are in the impermeable state. Typical fluorescence intensity profiles along 
the diameter of the capsules are presented in Fig.3 (bottom). We note that 
the level of fluorescence from the wall is higher than from the interior, 
which could be connected with some adsorption of the inner 
polyelectrolyte~\cite{17}. 

\begin{figure}[htbp]
\centerline{\includegraphics[width=6cm]{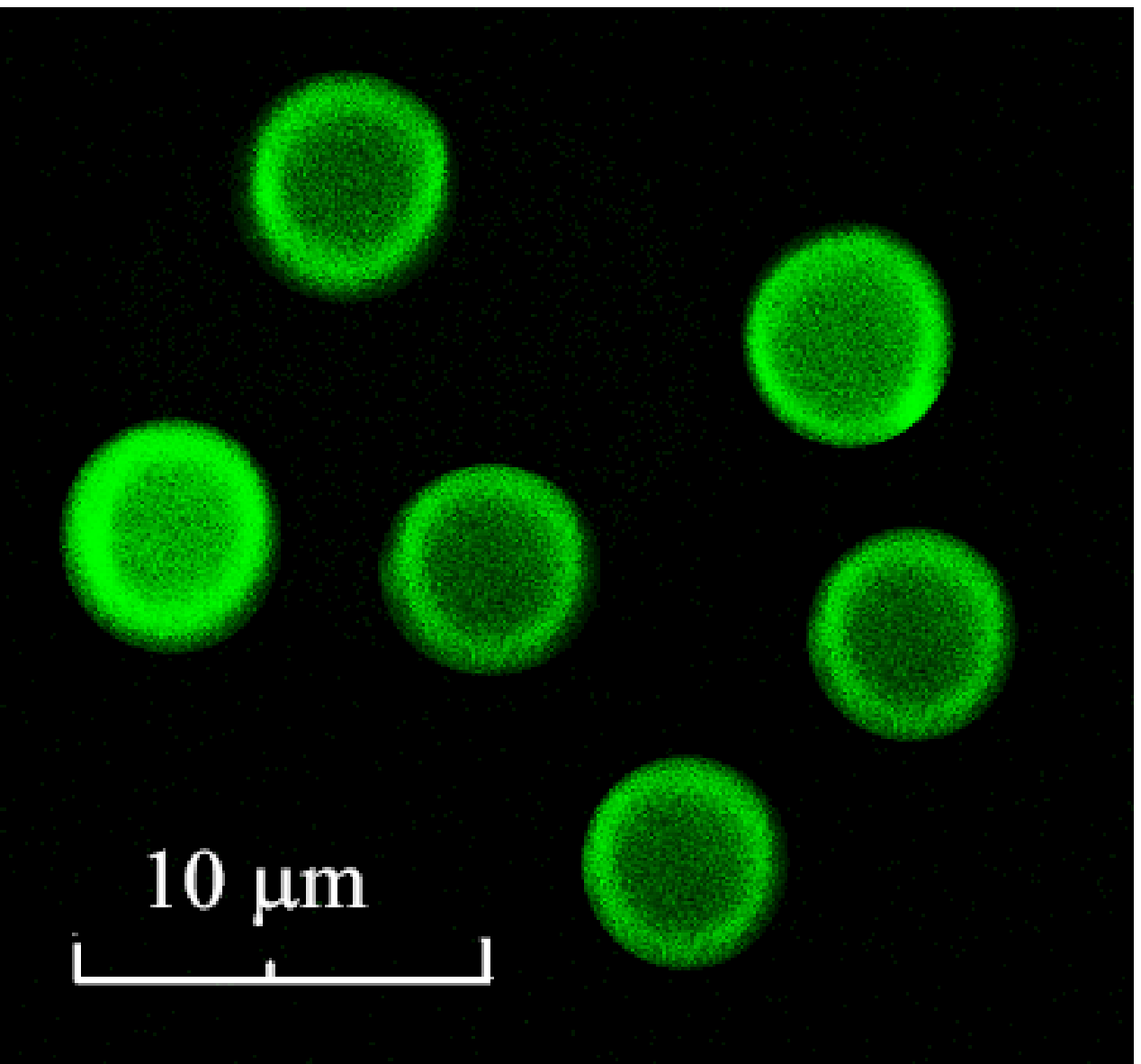}}
\centerline{\includegraphics[width=8cm]{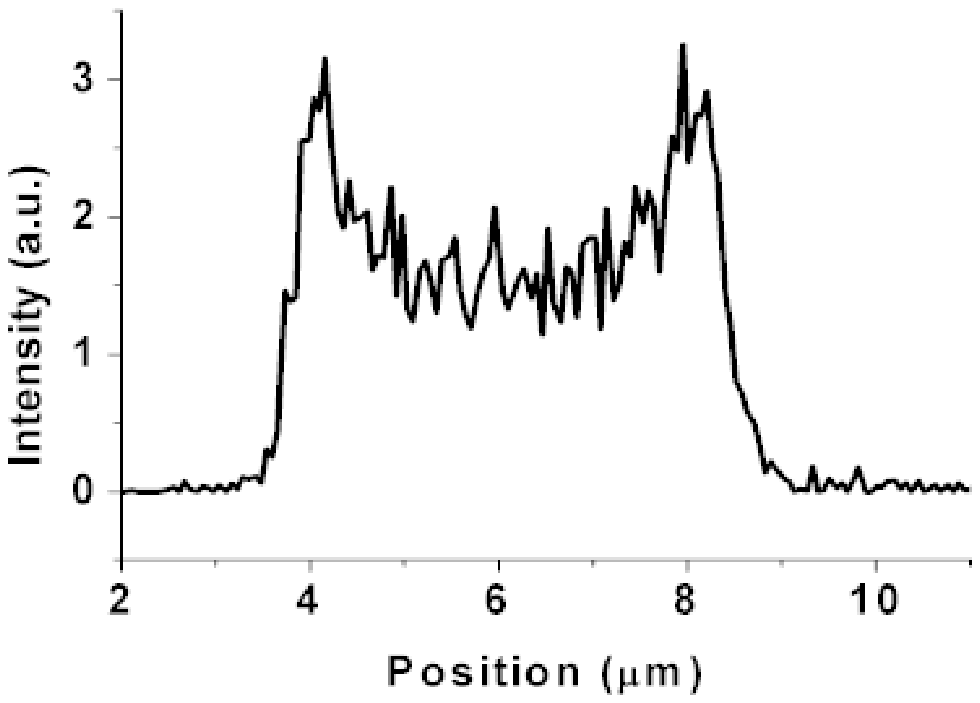}}
\caption{Confocal images of polyelectrolyte microcapsules filled with PSS (top) and typical fluorescence profile along the diameter of the capsule (bottom)}
\label{fig3}
\end{figure}

The size of the swollen capsules was determined as an average of 6-10 
capsules. The variability in size of the similarly prepared capsules was 
always within the error of optical measurements. We found that the 
``filled'' capsules are always larger than the original colloidal template, 
and that their radius depends on the size of the original template, the 
number of the bilayers in the shell, and the amount of encapsulated 
polyelectrolyte. These observations are consistent with our theoretical 
model. 

Fig.4 illustrates the typical dependence of equilibrium radius on the shell 
thickness. Here, the results for the first type of ``filled'' capsules 
(MnCO$_{3 }$ template) filled with PSS solutions of different concentrations, 
and made on the particles of the same size $r_0 = 1.85\pm 0.2 \quad \mu $m are 
given. It was previously found~\cite{3,28} that the thickness of 
a PSS/PAH bilayer is in the range of 3-5 nm. Here to evaluate the thickness 
of the shell with a known number of PSS/PAH bilayers we use the average 
value of 4 nm, as before~\cite{5,7}. The radius of the swollen 
capsules decreases with the shell thickness and is larger for the capsules 
with higher concentration of the inner polyelectrolyte solution. We fitted 
these experimental results to Eq. (\ref{eq6}) taking the combination of Young's 
modulus and the osmotic coefficient $E / \varphi $ as a fitting parameter 
and obtained the value of $E / \varphi \sim 200$ MPa for both curves 
presented in Fig.4. One can see that the predictions of the model are indeed 
confirmed by experiment, and the continuum mechanics approach is applicable 
for a molecularly thin multilayer film.

\begin{figure}[htbp]
\centerline{\includegraphics[width=8cm]{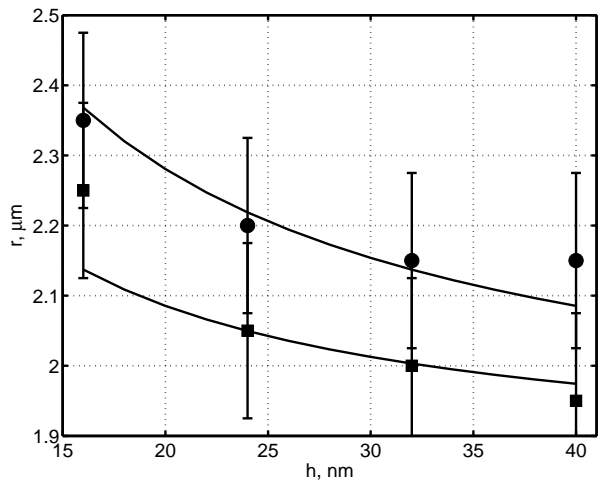}}
\caption{Radius of the swollen capsule made on the MnCO$_{3}$ template of radius 1.85$\mu $m as a function of the shell thickness. Two concentrations are shown: $0.16$ mol/L (squares) and $0.32$mol/L (circles). Fitting (solid curves) 
corresponds to $E / \varphi \sim 200\;\mbox{MPa}$ }
\label{fig5}
\end{figure}

\begin{figure}[htbp]
\centerline{\includegraphics[width=8cm]{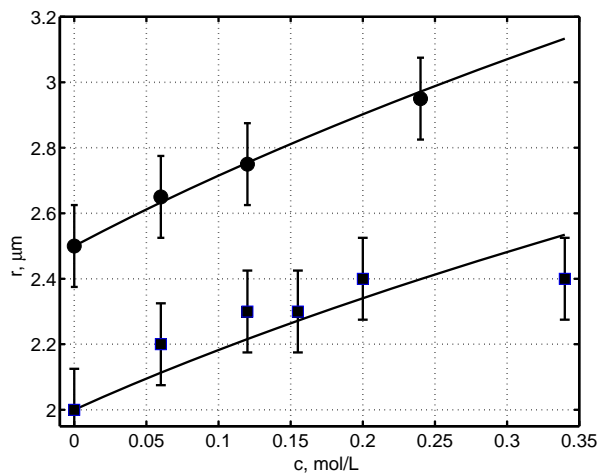}}
\caption{Radius of the swollen capsule as a function of monomer concentration in the inner solution. Fitting (solid curves) $E / \varphi \sim 300$ MPa (MnCO$_{3 }$template, circles) and $E / \varphi \sim 200$ MPa (MF template, squares)}
\label{fig6}
\end{figure}

The dependence of the equilibrium radius on the concentration of the inner 
polyelectrolyte, both the fitting curves and the experimental data, are 
shown in Fig.5. Here we present data both for the capsules of the first 
(MnCO$_{3 }$ template) and the second (MF template) types. The capsules are 
made on the templates of different size $r_0 = 2.5\pm 0.2\mu $m and $r_0 = 
2.0\pm 0.1 \quad \mu $m, correspondingly. The shell was always composed by four 
PSS/PAH bilayers. From the fit of experimental data we have obtained $E / 
\varphi \sim 300$ MPa for the first type, and $E / \varphi \sim 200$ MPa for 
the second type of capsules. 

One can conclude that, taking into account realistic values of the osmotic 
coefficients~\cite{15,17} for PSS solutions, Young's modulus $E$ found in 
our swelling experiment is of the order of 100 Mpa. This value is confined 
between the values found in the osmotic buckling experiment~\cite{5} 
and recent AFM results~\cite{7} and requires further comments. 

We remark and stress that current study gives the \underline {upper} 
possible value of Young's modulus. In reality, the excess osmotic pressure 
could be smaller than estimated from the known concentration of 
polyelectrolyte chains. One reason for such a decrease could be connected 
with a partial dissociation of the PSS, while our model assumed that it is 
fully dissociated. One also cannot exclude that some of the counter-ions 
might condense on the polyelectrolyte shell. Besides that, a portion of 
counter-ions could escape from the inner polyelectrolyte solution to the 
outer solvent, reducing the osmotic pressure difference. Such a possibility 
was not included in our model. All these effects (also ignored in 
Ref.~\cite{5}) will effectively lead to smaller values of the Young's 
modulus. 

Thus, our results strongly support the results of previous AFM 
studies~\cite{7} implying that we are dealing with an 
\textit{elastomer}~\cite{6}. In other words, the value of Young's modulus of the 
polyelectrolyte multilayer falls in the range characteristic for 
cross-linked rubbers. Such a mechanical behavior is probably due to strong 
interaction between polyanions and polycations in the multilayer. Our model 
also shows that the final permeability of the shell to water is very 
important for the stabilization of the microcapsules, increasing the 
threshold of the buckling transition~\cite{5}. 

We also note that the current method includes only a few assumptions, 
compared to the buckling transition and AFM measurements. We assume that the 
capsule deformations are small and elastic. These assumptions are also 
present in the theory of buckling and in the model used to describe capsules 
deformed in the AFM experiment. The theory of buckling transition assumes 
that the capsule is highly permeable to water even on a short time scale of 
the buckling deformation. The AFM experiment-based model postulates a priori 
a spherical shape of the capsule (except in the contact regions), 
conservation of capsule volume, and neglects bending deformations. There is 
no need for these assumptions in describing the capsule swelling, since the 
capsule shape is always spherical and the process of swelling is a slow 
process, i.e. the diffusion of water through the capsule shell takes place 
and does not affect the capsule swelling. 

\section{Conclusion}

We have provided a theoretical model, which relates Young's modulus, shell 
thickness, capsule radius and concentration of the inner polyelectrolyte 
solution. The validity of the model was experimentally verified, and 
confirms the applicability of the macroscopic continuum mechanics approach 
to polyelectrolyte multilayer microcapsules. Young's modulus of the 
molecularly thin polyelectrolyte multilayer was found to be of the order of 
100 MPa. This represents the \underline {upper }possible value for Young's 
modulus and is close to the elasticity of an \textit{elastomer}. Such a value reflects a high 
degree of local interactions between polyanions and polycations. Hence, 
mechanically at small deformations the polyelectrolyte multilayer resembles 
a cross-linked rubber material.

\section{Acknowledgements}

We acknowledge the support of the Alexander von Humboldt Foundation through 
a research fellowship (DA) and a Sofja Kovalevskaja award (GBS). We are 
grateful to I.L.Radtchenko for help with polyelectrolyte encapsulation, and 
to Y.Fedutik for MnCO$_{3}$ synthesis. M.Deserno, C.Holm, and V.Khrenov are 
thanked for helpful discussions. We also thank R.G.Horn for valuable remarks 
on the manuscript.

\end{document}